# Accurate estimates of asymptotic indices via fractional calculus

Sharmistha Dhatt[#] and Kamal Bhattacharyya[*]

Department of Chemistry, University of Calcutta, Kolkata 700 009, India

**Abstract**

We devise a three-parameter random search strategy to obtain accurate estimates of the large-coupling amplitude and exponent of an observable from its divergent Taylor expansion, known to some desired order. The endeavor exploits the power of fractional calculus, aided by an auxiliary series and subsequent construction of Padé approximants. Pilot calculations on the ground-state energy perturbation series of the octic anharmonic oscillator reveal the spectacular performance.

## 1. Introduction

Fractional calculus (FC) has received considerable attention over the last few decades [1] in a variety of situations. In the context of phase transitions involving power series with a finite radius of convergence, FC can be implemented [2] to obtain improved estimates of critical indices. We have recently showed [3] how it can be employed fruitfully in assessing asymptotic indices too from power series with zero radius of convergence. Here, we put forward a remarkably powerful strategy that exploits the idea of embedding an auxiliary series in conjunction.

From a Taylor series expansion of an observable $F(x)$,

$$F(x) = \sum_j f_j x^j, \quad x \to 0, \tag{1}$$

it is often of interest to extract its asymptotic ($x \to \infty$) behavior. Usually, a power-law form is assumed, much like the one around the critical point. This implies, the parameters $\alpha_0$ and $\beta_0$, defined by the equation

$$F(x) \simeq \alpha_0 x^{\beta_0}, \quad x \to \infty, \tag{2}$$

denote respectively the amplitude and exponent. Collectively, we call them the *asymptotic indices*. The problem of extracting (2) from (1) is quite involved, but possesses a very general character. For example, Bender *et al* [4] had chosen a number of examples, with $\beta_0 = 0$ in (2), to explore how difficult it is to estimate $\alpha_0$ from variants of (1), with known $\{f_j\}$ up to a certain number of terms.

The power-law form (2) is, in cases, intuitively obvious [3], *e.g.*, when $F$ stands for an observable and $x$ is a tunable physical variable. In a few situations, however, form (2) represents the leading behavior of an asymptotic expansion whose structure is derivable from (1). Choice (2) is thus quite reasonable. However, severity of the problem of getting (2) from (1) intensifies with the divergence of the parent expansion (1). Here, we choose (1) as a Rayleigh-Schrödinger perturbation expansion for energy from which form (2) emerges via a scaling argument.

Perturbation series for the anharmonic oscillator Hamiltonian

$$H(\lambda) = -d^2/dx^2 + x^2 + \lambda x^{2M} = H_0 + \lambda V \tag{3}$$

is well-known [5 - 8]. Specifically, the ground-state energy series

$$E_0^M(\lambda) = \sum_{j=0}^{\infty} \varepsilon_{0j}^0(M)\lambda^j \tag{4}$$

reveals via Symanzik's scaling argument [6] that

$$E_0^M(\lambda) = \lambda^{1/(M+1)} \sum_{j=0}^{\infty} \varepsilon_{0j}^{\infty}(M)\lambda^{-2j/(M+1)}. \tag{5}$$

Thus, while (4) is equivalent to (1), the leading term in (5) corresponds to (2) with

$$\beta_0 = \lambda^{1/(M+1)}, \alpha_0 = \varepsilon_{00}^{\infty}. \tag{6}$$

Most widely studied problems involve $M = 2, 3$ and $4$. An increase in $M$ actually worsens drastically the divergent character of (4). This is evident from the known [5 - 8] asymptotic growth of the coefficients in (4) as

$$\lim_{j \to \infty} \varepsilon_{0j}^0(M) \sim [(M-1)j]! A_M^j \tag{7}$$

where $A_M$ is some $M$-dependent constant. Thus, with energy as an observable, here we notice that, if the system Hamiltonian is given by (3), one has ready results for $\alpha_0$ and $\beta_0$ in (2) as seen in (6). While here the exponent is known and the only problem is to determine $\alpha_0$, both of them need to be evaluated for a series like (1) with no reference to some Hamiltonian origin.

Divergence of expansion (1) is commonly encountered in calculations of $F(x)$ at some large $x$-value. The most popular technique is the construction of suitable Padé approximants (PA) [7, 9]. In case of (4), however, straightforward performance of PA is poor, except for very small $\lambda$ and $M < 4$ [8]. In fact, a few effective variants of the PA in estimating values of $F(x)$ at large $x$ have recently been put forward [10] where major references to earlier works may be found. We also tried to get reasonable estimates of $\beta_0$ [11] and $\alpha_0$ [12] by using specific variants of the PA. A different sort of approach to obtain $F(x)$ at large $x$ is to employ multi-valued algebraic approximants [13] that are constructed in the same spirit as the PA. On the other hand, quite a few very successful methods of deriving strong-coupling expansions from weak-coupling ones for (3) have also come up from time to time (see, e.g., [14 -15] and refs. therein). However, they commonly rely on the large-$j$ behavior of $f_j$ [e.g., (7)], along with the scaling relation (5). All such studies make it clear that the case $M = 4$ is the most notorious.

In short, thus, the venture of extracting (2) from (1) becomes most challenging for the octic anharmonic oscillator (OAO) problem. To proceed, therefore, we view (4) as a purely numerical series, with no reference to any Hamiltonian origin, so that one can disregard the scaling in (5) or the known $\beta_0$ in (6). Only, in calculating $\alpha_0$, we employ the known $\beta_0$ value because the plan is to check the efficacy of the endeavor, and, it is evident, a rougher input $\beta_0$ would only worsen the target value sought. To achieve our end, we take (1), couple it to a 2-parameter auxiliary series, import FC and adopt an appropriate PA strategy. The overall endeavor is finally cast in the form of a 3-parameter random search problem. For computational purposes, we employ the coefficients $\varepsilon_{0j}^0 (M = 4)$ in (4) up to $j = 50$ [16].

**2. The strategy**

Let us start with an auxiliary series $A(x)$ that admits of both the forms (1) and (2). Simplest is to take

$$A(x) = (1 + px)^q, \tag{8}$$

with two variable parameters $p$ and $q$. We then use (1) to construct

$$H(x) = F(x) / A(x). \tag{9}$$

The asymptotic parameters of $H(x)$ will be

$$\alpha_0(H) = \alpha_0(F) / p^q, \quad \beta_0(H) = \beta_0(F) - q. \tag{10}$$

Likewise, the same parameters for $H'(x) = dH(x)/dx$ will turn out to be

$$\alpha_0(H') = \alpha_0(H)\beta_0(H), \quad \beta_0(H') = \beta_0(H) - 1. \tag{11}$$

We first delineate the plan of estimating $\beta_0(F)$. This requires elimination of $\alpha_0(F)$. So, we define a new series as

$$B(x) = xH'(x)/H(x). \tag{12}$$

It satisfies

$$\lim_{x \to \infty} B(x) = \beta_0(F) - q. \tag{13}$$

Therefore, sequences of diagonal PA to $B(x)$ can be evaluated in the $x \to \infty$ limit. These sequences are of the form

$$S_1^N = \left([N/N]B(x)\right)_{x=\infty}. \tag{14}$$

The limit point of such a sequence should converge to $\beta_0(F) - q$. To improve the convergence of $\{S_1^N\}$, we exploit FC in the following way. The Riemann-Liouville convention [1] allows us to define a fractional order ($g$) differential as

$$D^g y^n = \frac{\Gamma(n+1)}{\Gamma(n+1-g)} y^{n-g}. \tag{15}$$

Prescription (15) may be used to construct a function $B_1(x)$ where

$$B_1(x) = x^g D^g B(x). \tag{16}$$

If $B(x)$ has the form

$$B(x) = \sum_j b_j x^j, \tag{17}$$

then $B_1(x)$ will look as

$$B_1(x) = \sum_j b_j \frac{\Gamma(j+1)}{\Gamma(j+1-g)} x^j, \quad x \to 0. \tag{18}$$

On the other hand, we also have from (16)

$$\lim_{x \to \infty} B_1(x) = \tfrac{\Gamma(1)}{\Gamma(1-g)}\left(\beta_0(F) - q\right). \tag{19}$$

This implies, the PA sequences

$$S_2^N = \left([N/N]B_1(x)\right)_{x=\infty} \tag{20}$$

may be estimated using (18) and employed for the left part of (19) to yield a sequence of approximants for the true $\beta_0(F)$:

$$\beta_0^N(F) = \Gamma(1-g) S_2^N + q. \tag{21}$$

Result (21), however, depends critically on the chosen $p$, $q$ and $g$ that have been employed in (8) and (18). Thus, we require a final 3-parameter optimization step only, to which we shall return in due course.

Let us outline now the procedure for the evaluation of $\alpha_0(F)$. Here, we assume that $\beta_0(F)$ is somehow known *a priori*. We indeed choose

$$q = \beta_0(F). \tag{22}$$

As a result, the asymptotic parameters of $H(x)$ become

$$\alpha_0(H) = \alpha_0(F)/p^{\beta_0(F)}, \quad \beta_0(H) = 0. \tag{23}$$

Therefore, sequences of the form

$$S_3^N = ([N/N]H(x))_{x=\infty} \tag{24}$$

should hopefully converge to $\alpha_0(F)/p^q$, with $q$ given by (22). Importing FC, an added flexibility through fractional order differential $g$, defined in (15), can be instilled in the same way as has been done in going from $B(x)$ to $B_1(x)$. We define $H_1(x)$ as

$$H_1(x) = x^g D^g H(x). \tag{25}$$

and construct the PA sequences

$$S_4^N = ([N/N]H_1(x))_{x=\infty} \tag{26}$$

to obtain a sequence of gradually improved estimates for the amplitude $\alpha_0(F)$:

$$\alpha_0^N(F) = \Gamma(1-g)S_4^N p^{\beta_0(F)}. \tag{27}$$

Note that here a 2-parameter optimization step is involved.

Finally, we define an appropriate error and proceed to minimize it with respect to the parameters $p$, $q$ and $g$. This forms the optimization strategy. In case of $\alpha_0(F)$, $q$ is assumed known [see (22)], but can be replaced by the already evaluated $\beta_0^N(F)$ via (21) at each step of estimation of $\alpha_0^N(F)$ in (27). In either approach, therefore, the amplitude evaluation is a 2-parameter problem. Now, if we employ either (21) or (27) and use the coefficients up to $j = K$ in (1), the maximum number of diagonal PA that we can construct is $[L/L]$ where $L = K/2$. Thus, $N$ starts from 1 and continues up to $L$. Therefore, we find it convenient to measure the error $\Delta$ by

$$\Delta_\delta = \frac{1}{L-1}\sum_{N=1}^{L-1}\left(\frac{\delta_0^N}{\delta_0^L}-1\right)^2 \tag{28}$$

where $\delta$ stands for either $\beta$ or $\alpha$. This error desirably reveals that, for a faster converging sequence, it would reduce more in magnitude.

## 3. Results and discussion

The computational scheme proceeds as follows. We choose a specific property, *e.g.*, $\beta$ or $\alpha$. In the former case, we fix the three parameters (*p*, *q* and *g*) and *K*, the maximum number of coefficients in (1) to be used, to obtain the sequence of values via (21) and estimate the error $\Delta_\beta$ in (28). Similarly, the error $\Delta_\alpha$ is estimated at fixed *p*, *g* and *K*, only keeping *q* constant at (22). Then, via a random search strategy, we try to minimize $\Delta_\delta$. For sufficiently small errors, it has, however, been observed that very small changes in *p*, *q* and *g* (or, *p* and *g* only, for $\alpha$) may lead to comparable errors. So, we have fixed a sufficiently small error-level and take about 5000 different values of the variables, and their corresponding errors that are less than the pre-assigned $\Delta$. The average estimate for each of the parameters is then computed, along with the error.

Table 1 shows sample data for the exponent. The last entry is actually again an average estimate for $\beta_0^L(F)$ in (21) with the corresponding last PA ($L = K/2$) over all the values within the error-level under consideration. We note a smooth passage of $\beta_0^L(F)$ towards exactness (1/5) with increasing *L*.

In Table 2, we display the behavior of 3 specific sequences at their respective *average* values of *p*, *q* and *g*. It brings to light how good is the convergence of (21) and how results gradually improve with increasing input information. It also exhibits the rapidity with which values of the sequences of approximations settle as *K* is increased. At *K* = 30, the value remains constant over the region of *N* = 8 to *N* = 15 while the same constancy at a better value is found over *N* = 4 to *N* = 25 when *K* is raised to 50.

Similar feat is experienced in case of studying scheme (27) for the amplitude. Results are presented in Table 3. The last entry again furnishes the average $\alpha_0^L(F)$ in (27) where $L = K/2$ and generated sequences lie within the error-level under consideration. We note happily that these values of $\alpha_0^L(F)$ approach gradually the true estimate (1.22582) [5] as *L* grows.

In Table 4, the behavior of 3 different sequences of $\alpha_0^N(F)$ at their respective *average* values of *p* and *g* are shown. We note again both the rapidity of convergence of (27) and improvement in the estimates with increasing *K*. The sequences of approximations settle more quickly at higher *K*. However, now the range over which constancy is exhibited has been reduced compared with the same in the exponent case, for a fixed *K*. This only implies, computed amplitudes are somewhat inferior in quality.

We finally highlight the advantages of the FC-based strategies. Table 5 shows the chief gains in brief. All displayed data are rounded off at the 3$^{rd}$ decimal place. Here, method I refers to the parent scheme and method II the bare FC-assisted one [3]. Notably, the latter involves a 1-parameter variation, the order of fractional differentiation. Comparing such results with those of method III, the 3-parameter variational route proposed in this work, we see that the gains are indeed dramatic. Use of just 10 perturbation coefficients in method III leads us to quite reliable data, far superior to what one can have by taking K = 50 and adopting either of the first two methods.

**4. Concluding remarks**

To summarize, we have found a very efficient strategy for the calculations of asymptotic indices via FC by introducing the idea of an auxiliary series. The gain is spectacular for the OAO problem, if we remember the earlier estimates [3]. The percentage error reduces by more than an order of magnitude at a given *K* in going from method II to method III. More strikingly, method III offers far better results at *K* = 10 than what method II can yield even at *K* = 50. Such dramatic improvements could not be appreciated for *M* < 4 cases, and this is precisely why we have considered here the OAO problem.

**Acknowledgement**

SD wishes to thank CSIR, India for a fellowship.

Table 1. A comparative survey of the estimates for $\beta_0(F)$ in the OAO case at varying $K$. The error $\Delta_\beta$ refers to (28). Average values of the parameters and the error are displayed.

| $K$ | $p$ | $q$ | $g$ | $\Delta_\beta$ | $\beta_0^L(F)$ |
|---|---|---|---|---|---|
| 10 | 28.55490 | 0.19675 | 7.08512 | 9.53 E(-13) | 0.196941 |
| 20 | 28.23490 | 0.19725 | 8.22487 | 3.59 E(-12) | 0.197464 |
| 30 | 28.07506 | 0.19785 | 8.96494 | 1.16 E(-11) | 0.198073 |
| 40 | 27.90488 | 0.19904 | 9.88560 | 5.88 E(-11) | 0.199273 |
| 50 | 27.74501 | 0.19975 | 9.86507 | 8.30 E(-11) | 0.199979 |

Table 2. Behavior of $\beta_0^N(F)$ in (21) for the OAO case at 3 different $K$-values.

| $N$ | $K = 10$ | $K = 30$ | $K = 50$ |
|---|---|---|---|
| 1 | 0.1969370 | 0.1980724 | 0.1999810 |
| 2 | 0.1969372 | 0.1980726 | 0.1999812 |
| 3 | 0.1969375 | 0.1980738 | 0.1999829 |
| 4 | 0.1969372 | 0.1980744 | 0.1999812 |
| 5 | 0.1969372 | 0.1980738 | 0.1999812 |
| 6 |  | 0.1980744 | 0.1999812 |
| 7 |  | 0.1980729 | 0.1999812 |
| 8 |  | 0.1980744 | 0.1999812 |
| 9 |  | 0.1980744 | 0.1999812 |
| 10 |  | 0.1980744 | 0.1999812 |
| 15 |  | 0.1980744 | 0.1999812 |
| 25 |  |  | 0.1999812 |

Table 3. A comparative survey of the estimates for $\alpha_0(F)$ in the OAO case at varying $K$. The error $\Delta_\alpha$ refers to (28). Average values of the parameters and the error are displayed.

| $K$ | $p$ | $g$ | $\Delta_\alpha$ | $\alpha_0^L(F)$ |
|---|---|---|---|---|
| 10 | 2.97497 | 1.73091 | 8.07 E(-8) | 1.23528 |
| 20 | 2.96850 | 1.68849 | 1.77 E(-7) | 1.23222 |
| 30 | 2.96501 | 1.66513 | 2.14 E(-7) | 1.23035 |
| 40 | 2.96050 | 1.65051 | 2.25 E(-7) | 1.22888 |
| 50 | 2.93490 | 1.64599 | 2.29 E(-7) | 1.22630 |

Table 4. Behavior of $\alpha_0^N(F)$ in (27) for the OAO case at 3 different $K$-values.

| $N$ | $K = 10$ | $K = 30$ | $K = 50$ |
|---|---|---|---|
| 1 | 1.2359381 | 1.2322716 | 1.2288046 |
| 2 | 1.2354543 | 1.2309072 | 1.2271180 |
| 3 | 1.2354504 | 1.2306526 | 1.2267297 |
| 4 | 1.2354094 | 1.2306091 | 1.2266294 |
| 5 | 1.2353001 | 1.2306063 | 1.2266061 |
| 6 |  | 1.2306091 | 1.2266294 |
| 7 |  | 1.2306092 | 1.2266338 |
| 8 |  | 1.2306091 | 1.2266294 |
| 9 |  | 1.2306108 | 1.2266294 |
| 10 |  | 1.2306091 | 1.2266294 |
| 15 |  | 1.2306091 | 1.2266294 |
| 25 |  |  | 1.2266294 |

Table 5. A comparative survey of the estimates of asymptotic parameters for the OAO case by different methods (see text) at varying $K$.

| Property | Exact value | Method | $K$ | | |
|---|---|---|---|---|---|
|  |  |  | 10 | 30 | 50 |
| Exponent ($\beta$) | 1/5 | 1 | 0.035 | 0.042 | 0.044 |
|  |  | 2 | 0.125 | 0.135 | 0.138 |
|  |  | 3 | 0.197 | 0.198 | 0.200 |
| Amplitude ($\alpha$) | 1.226 | 1 | 1.865 | 1.822 | 1.816 |
|  |  | 2 | 0.785 | 0.857 | 0.877 |
|  |  | 3 | 1.235 | 1.230 | 1.226 |